\documentclass[11pt]{article}
\begin{document}
\begin{center}
{\large\scshape

Source to Accretion Disk Tilt
}
\vspace{3mm}
\end{center}

\begin{center}
{  
M.M. Montgomery$^{1}$ and E. L. Martin$^{1,2}$
}
\end{center}

\begin{center}
\noindent
{\small \em
$^{1}$ Department of Physics, University of Central Florida, Orlando, FL  32816, USA\\
        $^{2}$ Centro de Astrobiologia CSIC - INTA, Madrid, Spain
}
\end{center}
{
\begin{abstract}
Many different system types retrogradely precess, and retrograde precession could be from a tidal torque by the secondary on a misaligned accretion disk.  However, a source to cause and maintain disk tilt is unknown.  In this work, we show that accretion disks can tilt due to a force called lift.  Lift results from differing gas stream supersonic speeds over and under an accretion disk.  Because lift acts at the disk's center of pressure, a torque is applied around a rotation axis passing through the disk's center of mass.  The disk responds to lift by pitching around the disk's line of nodes.  If the gas stream flow ebbs, then lift also ebbs and the disk attempts to return to its original orientation.  

To first approximation, lift does not depend on magnetic fields or radiation sources but does depend on mass and the surface area of the disk.  Also, for disk tilt to be initiated, a minimum mass transfer rate must be exceeded.  For example, a $10^{-11}M_{\odot}$ disk around a 0.8$M_{\odot}$ compact central object requires a mass transfer rate greater than $\sim10^{-13}$M$_{\odot}$yr$^{-1}$, a value well below known mass transfer rates in Cataclysmic Variable Dwarf Novae systems that retrogradely precess and that exhibit negative superhumps in their light curves and a value well below mass transfer rates in protostellar forming systems.  
\end{abstract}


\section{Introduction}
Over the last forty years or so, cyclic brightness modulations that have periods longer than orbital periods have been observed in a variety of systems including Cataclysmic Variables (CVs) such as AM CVn, V503 Cyg, V795 Her, and TT Ari (see Montgomery 2009a,b and references within); Active Galactic Nuclei (AGN) and black hole binaries such as SS 433 (see Foulkes et al. 2010 and references within); X-ray binaries (XB) such as V1405 Aql (Retter et al. 2002) which are X-ray sources analogous to Her X-1 (Tananbaum et al. 1972) and LMC X-4 (Ilovaisky et al. 1984).  These long periods are referred to as retrograde precessional periods or super-orbital periods.  In these systems, the accretion disk is thought to be retrogradely precessing.  Other observations such as jet retrograde precession and changes in jet direction from jet producing sources like AGNs, X-ray binaries, and young stellar objects (YSOs) are also thought to be caused by the retrograde precession of their underlying disks (Lai 2003).  Evidence comes from observations:   The systematic velocity variations of the optical jet emission, the radio jet morphology, and optical photometry in SS 433 (Maloney \& Begelman 1997a and references within) indicate the same retrograde precessional period.  As the jet is believed to originate from the inner disk, photometry is from the outer disk, and disks are expected to precess at the same rate, jet precession could be another observational beacon to disk retrograde precession.  

Maloney \& Begelman (1997a) suggest that retrogradely precessing, inclined accretion disks may be common in X-ray binaries.  Similarly, Warner (2003) suggests a common phenomenon occurs in X-ray binaries and CVs to explain retrograde precession.  Warner (2003) also suggests that long period modulations are taken to be indirect evidence for a tilted and precessing disk in these systems.  In Montgomery (2009a), we show that a source to retrograde precession in black holes, neutron stars, pulsars, X-ray binaries, Cataclysmic Variables, AGNs, protoplanetary and protostellar systems could be the net tidal torque by the secondary on a misaligned accretion disk, like the net tidal torque by the Moon and the Sun on the equatorial bulge of the spinning and tilted Earth.  Disk tilt and retrograde precession appear to be common to many types of systems that have accretion disks.  However, a common source to disk tilt remains a mystery:  Katz (1973) imposes a disk tilt boundary condition to produce a precessing tilted disk but does not suggest a source to disk tilt.  Smak (2009) suggests that maintenance of disk tilt could be from a stream-disk interaction, but questions the source to disk tilt and the source to transition a disk from tilted to coplanar.  In this work, we attempt to answer these two questions.  

Accretion disks have been postulated to warp or tilt via a potpourri of sources.  For example, disk tilt in X-ray binaries can be from gas streaming at an upward angle from the inner Lagrange point for one half of the orbit and at a downward angle for the second half of the orbit (Boynton et al. 1980).  In some CVs, a disk tilt can be held constant by a gas stream that is fed via the magnetic field of the secondary (Barrett, O'Donoghue, \& Warner 1988).  Also for CV systems, a disk tilt instability can result from a coupling of an eccentric instability to Lindblad resonances (Lubow 1992).  For X-ray binaries and YSOs, tidal force from the companion star could drive the precession (e.g., Katz 1973, Wijers \& Pringle 1999, Terquem et al. 1999, Bate et al. 2000), Ogilvie \& Dubus 2001).  A vertical resonant oscillation of the disk mid-plane can be caused by tidal interactions between a massive secondary and a coplanar primary (Lubow \& Pringle 1993).  A warping instability could be caused by irradiation from the primary (Pringle 1996, 1997).  A warping could aso be caused by direct tidal forces from a secondary orbiting on an inclined orbit (e.g., Clarke \& Pringle 1993, Hall et al. 1996, Papaloizou \& Terquem 1995, Larwood et al. 1996, Larwood 1997, Larwood \& Papaloizou 1997).  A disk tilt can be induced if the secondary has an axis that is tilted relative to the orbital plane (Roberts 1974).  A disk warp can be caused by misalignments of the spin axis of a compact and/or magnetized primary and the disk axis (see e.g., Kumar 1986, 1989, Lubow \& Pringle 2010 for protostellar disks).  To warp a disk, Foulkes et al. (2010) impose a requirement of steady nuclear burning on the surface of the white dwarf of CV KR Aur.  Another method invoked to warp a disk is using an asymmetric coronal disk wind force on the disk surface (Schandl \& Meyer 1994, Schandl 1996) which is applicable to X-ray emitting neutron stars.  Quillen (2001) suggests a warp could be caused by a wind blowing across the face of a disk from AGNs and other objects that generate energetic outflows.  Hobbs \& Nayakshin (2009) suggest that infall of gas clouds on inclined orbits can possibly warp the central parsec around the Galactic Center.  For accretion onto magnetic (neutron, white dwarf, T Tauri) stars, many authors (e.g., Aly 1980, Lipunov \& Shakura 1980, Lai 1999, Terquem \& Papaloizou 2000) suggest that the magnetic field can induce disk warping and precession to explain quasi-periodic oscillations in low-mass X-ray binaries, millihertz variability in accreting X-ray pulsars, and irregular variabilities in T Tauri stars.  For galactic warp, Jiang \& Binney (1999) show that accretion reorients the halo angular momentum vector from the original symmetry axis of the galaxy, the halo becomes warped, and disk warp is the tracer to the halo-warp.    

Many of these models are limited to a few system types such as central radiation or magnetic sources and/or impose conditions such as magnetic fields, inclined orbits, steady nuclear burning on the surface of non-magnetic white dwarf, etc.  For precession to occur, the disk tilt needs to be excited and maintained. However in most accreting binary systems, the disk plane is believed to remain aligned with the orbital plane.  In addition, many of the models suffer flaws:  Roberts (1974) suggestion may generate a disk tilt, however, the tilt may decay by tidal damping on a timescale shorter than the circularization time (Chevalier 1976).  Radiation-driven warping instability may occur in XBs, however Ogilvie \& Dubus (2001) finds that this source does not explain all long-term XB variabilities.  Ogilvie \& Dubus (2001) also surmise that radiation-driven warping is probably not a common occurrence in low-mass X-ray binaries.  Maloney \& Begelman (1997a) note that torques by a companion star must dominate over radiation-driven warping.  Maloney et al. (1996) find that a high radiation efficiency is needed to explain the warp in some systems such as NGC 4258, and emission from the warped disk may not match the frequency and shape of the brightness modulations.  Although dragging of inertial frames by a rotating black hole can cause retrograde precession, this Lense \& Thirring (1918) technique is limited to very massive compact central objects that have non-colinear disk and black hole spin axes, and the success of this technique in generating a disk warp is only within a transition radius of the black hole (outside this radius, the disk does not feel the effect).  Although threading of accretion disks with curved magnetic fields may generate a warping instability that results in a disk retrograde precession (Lai 2003), the origin of the disk-threaded magnetic field is not well understood.  Murray \& Armitage (1998) find that CV DN accretion disks do not tilt significantly out of the orbital plane by vertical instabilities. As noted in Maloney \& Begelman (1997b), other suggested sources to disk tilt have difficulty communicating a single precession frequency through a differentially rotating fluid disk.  For example, in YSOs, the disk may initially be mis-aligned, however differential precession and internal stresses dissipate energy to damp disk warp (Lubow \& Ogilvie 2000).  

In this work, we search for a common source to generate and maintain an accretion disk tilt.  In \S 2, we develop theoretical expressions to generate and maintain a disk tilt and how and why accretion disks tilt.  In \S 3, we discuss some results and in \S 4, we provide a summary, conclusion, and future work.

\section[]{Theoretical Expressions for Disk Tilt}
\subsection{Torques and Forces on Tilted Disks}
In Montgomery (2009a) we introduce a connection between tidal torques by the Sun and the Moon on the oblate, spinning, tilted Earth and tidal torques  by secondaries on spinning, tilted accretion disks around primaries.  Both the Earth and the tilted disk can thus retrogradely precess by an applied average tidal torque:
\begin{eqnarray}
\bf{\bar{\Gamma}_{x}} & =  & \bf{r} \bf{\times} \bf{\bar{F}}  \nonumber \\
                                          & =  & \bf{r} \bf{\times} \bf{\bar{F}_{Rz}}  \nonumber \\
                                        & =  & |d\bf{\bar{F}_{Rz}}| \bf{\hat{x}} \nonumber \\
                                        & = & \left| \frac{3GM_{2}}{2d^{4}}(I_{z'z'}-I_{x'x'})\cos\theta \right| \bf{\hat{x}} \nonumber \\
                                        & = & \left| \frac{3GM_{2}M_{d}r_{d}^{2}}{8d^{4}}\cos\theta \right| \bf{\hat{x}}
\end{eqnarray}
\noindent
where, in cartesian coordinates, $\bf{\bar{F}}$ is any average force located a radial distance $\bf{r}$ from the origin such as $\bf{r}$=$d\bf{\hat{y}}$, $d$ is distance between the primary and secondary, $\bf{\bar{F}_{Rz}}$ is a component of gravitational force, $G$ is the universal gravitational constant, $M_{2}$ is the secondary mass of the binary, \( [(I_{z'z'}-I_{x'x'}) = I_{x'x'} = 1/4M_{d}r_{d}^{2}] \) are components of moment of inertia, $M_{d}$ is the mass of the disk, $r_{d}$ is the disk (assumed circular) radius, and $\theta$ is the obliquity angle.  Our Equation (1) is very similar to Equation (16) in Montgomery (2009a).  

Equation (1) shows that a gravitational torque by the secondary results in retrograde precession of a tilted disk around the $\bf{z}$ axis (see Figure 1 which is based on Figure 1 in Montgomery 2009a).   As shown in Figure 1, $\bf{(x, y, z)}$ are the orbital plane axes and $\bf{(x', y', z')}$ are the body axes of the disk ($\bf{y'}$ not shown).  Note that $\bf{x}$ and $\bf{x'}$ share the same axes and $\bf{z', z}$ and $\bf{y}$ are in the same plane.  Rotation of the disk is taken to be around its polar axis $\bf{z'}$ to eliminate free precession.  The magnitude of the rotational angular velocity around this polar axis is $\dot{\epsilon}$.  In the figure, the primary (not shown) is located at the origin, a point that coincides with the disk's center of mass.  The point-mass secondary $M_{2}$ is located a separation distance $d$ from the primary that is along the $\bf{y}$ axis.  The $\bf{z}$ axis defines the normal to the orbital plane, and the magnitude of the orbital motion $\dot{\omega}$ is prograde.  Obliquity is the angle between the $\bf{z}$ and the $\bf{z'}$ axes as well as the angle between -$\bf{F_{G}}$, which is in the $\bf{z',z,y}$ plane, and the $\bf{y}$ axis (bottom panel).  $\bf{F_{G}}$ is the gravitational force between the primary and a gas particle in the disk located a distance $\frac{3r_{d}}{4}$ from the origin.   In the top panel, $\bf{F_{L}}$ and $\bf{F_{net}}$ are equal and opposite forces in the $\bf{z',z,y}$ plane and are also located a distance $\frac{3r_{d}}{4}$ from the origin.  Figure 1 exaggerates distances, sizes of objects, and sizes of vectors to show details.  

In Equation (1), the mass of the disk is the sum of the mass of the primary and the mass of each gas particle $m$ in the disk.  As expected, most of the mass of the disk is contained in the primary.  However, the primary is located at the origin, a point that is on the axis of the disk's rotation.  Therefore, the disk's moment of inertia and angular momentum are mostly affected by the gas particles in the disk.  As the primary contributes little to disk's moment of inertia and angular momentum, the disk is more easily affected by outside forces.  For example, a change in the tilted disk's angular momentum vector over time can be due to the secondary's gravitational force torqueing the tilted disk (Montgomery 2009a), and the disk responds by retrogradely precessing around the $\bf{z}$ axis (we note that if the primary is also oblate and tilted, then it too will retrogradely precess due to gravitational torque, but this discussion is outside the scope of this work).  

Unless a source is available to maintain disk tilt, disk tilt can decay over time by tidal damping:  Primary and secondary gravitational forces on the disk could cause the disk to torque around the $\bf{x'}$ body axis.  For example, we consider an isolated gas particle in the disk that is located a distance (3/4)$r_{d}$ from the origin (lower panel of Figure 1).  The magnitude of gravitational force between the gas particle and the secondary is $ \bf{|F_{mM_{2}}|} $ = $GmM_{2}r_{mM_{2}}^{-2} = $$GmM_{2}(d^{2}+\frac{9}{16}r_{d}^{2}-\frac{3}{2}r_{d}d\cos\theta)^{-2}$.  Using geometry, the $\bf{z}$-component magnitude of this force is
\begin{equation}
 |F_{mM_{2_{z}}}|= \frac{3r_{d}GmM_{2}}{4(d^{2}+\frac{9}{16}r_{d}^{2}-\frac{3}{2}r_{d}d\cos\theta)^{3/2}} \sin\theta.
\end{equation}
\noindent
Similarly, the magnitude of the gravitational force between the same gas particle and the primary star is $|\bf{F_{mM_{1}}}|$ = $(16/9)GmM_{1}r_{d}^{-2}$.  Using geometry, the $\bf{z}$-component magnitude of this force is 
\begin{equation}
|F_{mM_{1_{z}}}| = \frac{16GmM_{1}}{9r_{d}^{2}} \sin\theta.
\end{equation}
\noindent
The net force $\bf{F_{net}}$ is the sum of all forces.  If internal forces sum to zero then we can assume \( \bf{|F_{net}| \approx |F_{mM_{1_{z}}}| + |F_{mM_{2_{z}}}}|\).  Note that if the obliquity is zero then so is $\bf{F_{net}}$, as expected.  

Note that $\bf{|F_{net}}|$ can be further simplified if gravitational effects by the secondary are minimal:  If the primary is the more massive star (e.g., black holes) or the secondary is located very far away, the more the primary should contribute to damping disk tilt or realigning the angular momentum vectors.  That is, if $M_{1} >> M_{2}$ or $d>>r_{d}$, then  \( \bf{|F_{net}| \approx |F_{mM_{1_{z}}}}| \).  For example, if $M_{1}$=0.8$M_{\odot}$, $M_{2}$=0.32M$_{\odot}$, m=2x10$^{14}$kg, $r_{d}$=0.43d, d=1.23$r_{\odot}$ (see Montgomery 2009b), and assuming small angles, then \(  |F_{mM_{2_{z}}}| \approx 4.4\times10^{-2} |F_{mM_{1_{z}}}| \) and the gravitational force of the primary strongly acts to restore the disk to the orbital plane.  

If no rotation is to occur around the $\bf{x'}$ axis, then $\Sigma\bf{F}$=0 from Newton's Second Law and \( \bf {|F_{L}| = |F_{net}}| \) (see Figure 1).  Because approximately half the mass of the disk is located between the primary and secondary, $|\bf{F_{L}}|$ has to act on approximately half the gas particles in the disk.  Therefore, 
\begin{eqnarray}
|F_{L_{1}}| & \approx & |F_{net}| \nonumber \\
                    & \approx & |F_{mM_{1_{z}}}| + |F_{mM_{2_{z}}}| \nonumber \\
                    & \approx & \frac{8G \Sigma m M_{1}}{9r_{d}^{2}} \sin\theta + \frac{3r_{d}GmM_{2}}{8(d^{2}+\frac{9}{16}r_{d}^{2}-\frac{3}{2}r_{d}d\cos\theta)^{3/2}} \sin\theta 
\end{eqnarray}
\noindent
where $\Sigma$m is the sum of gas particles in the disk (i.e., the total mass of the disk, not including the mass of the primary).  

When forces do not balance, $\Sigma\bf{F}$=$m$$\frac{d\bf{v}}{dt}$, where $\bf{v}$ is velocity and $t$ is time, and $\Sigma\bf{F}$ can cause a change in velocity or a change in direction.  Likewise, a change in velocity or direction can cause $\Sigma\bf{F}$.  For example, if a gas stream flowing under a disk (see e.g., Frank, King, \& Lasota 1987 and references within) has changed direction compared to the path of the gas stream flowing over the disk then $\Sigma\bf{F}$ is generated. By Newton's Third Law, an opposing force to $\Sigma\bf{F}$ is also generated, and the disk reacts by moving perpendicular to the path of the under flow until forces balance.  This opposing force is $\bf{F_{L_{1}}}$ in Equation (4) and is one variation (hence the subscript 1) of a force commonly called the lift force or lift for short.  Notice that, to first approximation, lift does not depend on magnetic fields or radiation sources.

\subsection{How Accretion Disks Tilt}
All accretion disks can tilt by lift if certain conditions are met (see discussion on why disks tilt below). Lift is a mechanical force and is generated when a fluid such as a gas makes contact with and interacts with a solid body.   For lift to be generated, motion between the object and the fluid is demanded.  Lift can occur when a solid object moves through a static fluid (e.g., commercial airplane moving through air) or, in our case, the fluid moves past a solid-like object (i.e., an accretion stream moving over and under an  accretion disk).  However, no motion leads to no lift in both examples. In our case, if the gas stream ceases (or ebbs) then lift ceases (or ebbs) and angular momentum vectors of the compact central object and disk attempt to realign and restore the disk to its original orientation. 

Lift acts on an object at a point called the center of pressure.  The center of pressure is approximately one-fourth the radial distance from the leading edge as found from experiments in the lab on airfoils.  Lacking experiments on location(s) of the center of pressure(s) in fluid accretion disk(s), we have to make assumptions.  Although an accretion disk is a fluid, it can respond like a solid body:  Papaloizou \& Terquem (1995) analytically show, and Larwood et al. (1996) numerically verify, that rigid body precession of a Keplerian disk is possible so long as the sound crossing time scale in the disk is small compared to the precession time scale of the disk.  In addition, numerical simulations by Murray (1998) and observations by Patterson et al. (1998) do not indicate that disks differentially precess.  Three dimensional Smoothed Particle Hydrodynamic numerical simulations of artificially tilted accretion disks in Montgomery (2009b) show that a fluid disk remains a unit while the tilted accretion disk wobbles through its retrograde precession.  An example of a face-on and edge-on disk from the same simulation run in Montgomery (2009b) is shown in Figure 2.  The left panel shows a face-on tilted disk in relatively density whereas the right panel shows a cut-away of an edge-on non-tilted disk in relative internal energy.  Note a higher density in the rim of the numerically simulated elliptical disk is expected (see e.g., Frank, King, \& Lasota 1987).  As a result of the above, we shall assume that the disk acts like a rigid body, the leading edge is defined by the bright spot, and the center of pressure is near (3/4)$r_{d}$.   

\subsection{Why Accretion Disks Tilt}
Factors that can initiate and otherwise affect dynamic lift include the a) shape, b) size, and c) mass of the object experiencing lift, the d) velocity and e) inclination of flow past the object experiencing lift, and the f) density, g) viscosity, and h) compressibility of the gas stream.  An in-depth analysis of each of these effects is beyond the scope of this work.  We will not discuss the viscosity (which would invoke discussions on magneto-hydrodynamics) and compressibility of the gas.  However, we shall discuss the others.  

Hydrodynamic lift on an object can be enhanced if the object has an asymmetric shape, a symmetric shape but asymmetric orientation, asymmetric surface texture, and spin.  Of these four, we eliminate three in this work:  We assume the accretion disk has a symmetric shape around its three body axes (although this may not be the case in certain circumstances like during disk prograde precession); we assume the surface texture of each disk face is the same; and we assume the disk is not spinning like a flying, spinning tennis ball (we also assume gas particles do not spin like flying, spinning tennis balls either).  Because the accretion disk has an asymmetrical orientation with respect to the gas stream (see e.g., Lubow \& Shu 1975), an angle of attack is created between the gas stream and the disk rim.  Because of this angle of attack, because the gas particles in the disk rim are moving in the same direction and obliquely relative to the gas stream, and because the disk rim is not flat but does vary in height to and from the bright spot (see Figure 2 and discussions in Hartmann et al. 1998), lift can be enhanced.  Lift enhanced by disk shape can further cause the gas stream to change its flow path and speed, thereby further enhancing the disk tilt.  We should note that the shape of the accretion disk relative to the gas flow is very streamlined and hydrodynamic thereby reducing drag.  Drag is another force that acts on the disk but its direction is opposite that of the gas stream.  Like lift, drag forces are usually determined in a laboratory as these forces are based on the density and composition of the gas stream as well as the shape of the object experiencing lift, for example.  We shall minimize discussions of drag in this work so that we may focus on lift.  

The size and mass of the disk also affect lift as noted in Equation (4).  From experimental studies on airplanes, lift is mostly generated in surface area of the wings.  If the surface area is doubled, then the lift force is also doubled, hence the reason why airplanes have large wings.  As shown in Figure 2, an accretion disk is expected to have an elliptical shape and hence it has a large wingspan (i.e., elliptical disk's semi-major axis) relative to its depth (i.e., elliptical disk's semi-minor axis).  We should also point out that air flows over and under the entire surface area of the wing thereby affecting the entire wing.  In accretion disks,  the gas stream is narrow and thus flows over and under only a radial portion of the disk.  However, because fluid motion in accretion disks is expected to be Keplerian, Keplerian motion is faster than the orbital motion, and particles in the disk move under the gas stream that has overflowed the disk rim at the bright spot, the entire disk seems to be affected by the overflowing gas stream. 

The velocity and inclination of the accretion stream flow past an object can initiate lift although an analytical expression is more difficult to obtain precisely.  However, we can generate a simple analytical analysis to predict a gross magnitude of the lift.  As shown in Figure 3, the variation in the spacing of streamlines over and under the disk indicate differing amounts of pressure on the disk.  The wider the spacing of the stream lines, the speed of the gas is slower, and the pressure is greater.  By definition, pressure is force per unit area and the force associated with the pressure is lift.  The pattern of streamlines shown in Figure 3 is consistent with Bernoulli's equation, an equation that is a reformulation of the conservation of mechanical energy but applied to fluid flow.  A simplified variation of Bernoulli's equation to over ($o$) and under ($u$) the the accretion disk is
\begin{eqnarray}
P_{o} + \frac{1}{2}\rho v_{o}^{2} + \rho |a_{G}| r_{o} & = & P_{u} + \frac{1}{2}\rho v_{u}^{2} + \rho |a_{G}| r_{u} \\
(P_{u} - P_{o}) & = & \frac{1}{2} \rho (v_{o}^{2} - v_{u}^{2}) + \rho |a_{G}| (r_{o}-r_{u}) \\
|F_{L_{2}}| & \approx & \frac{A_{s}}{2} \rho (v_{o}^{2} - v_{u}^{2}) \\
|F_{L_{2}}| & \approx & \frac{\dot{M} |v_{o}| }{4}  \left( \frac{r_{d}}{b}\right)^{2} (1-\beta^{2}).
\end{eqnarray}

\noindent
In these equations, $P$ is pressure, $\rho$ is density, $\bf{|v|}$ is magnitude of velocity, $|\bf{a_{G}}|$ is the magnitude of the acceleration due to the universal gravitational constant $G$, $r$ is the distance from the origin to a gas particle in the disk, $\bf{|r_{o} - r_{u}|}$ = $\alpha H$, $\alpha$ is a positive fraction of the disk height $H$, $A_{s}=\pi r_{d}^{2}$ is a conservative estimate of the surface area of the disk that is between the primary and the gas stream, $\dot{M}$ is the gas stream mass transfer rate, and $b$ is the radius of the gas stream just prior to striking the bright spot.

The second term on the right hand side of Equation (6) is negligible if ${r_{o} \approx r_{u}}$ or if the second term is small compared to the first.  Distances from the origin to the top and bottom of the disk rim at the bright spot are likely to be similar in accreting systems and disk radii large enough that $|a_{G}|$ applies nearly equally to the top and bottom of the disk at the rim.  We do note that if \( (r_{o}-r_{u}) \sim H \), then the second term is likely to not be small as $|\bf{a_{G}}|$ is not small.  In this work, we assume ${r_{o} \approx r_{u}}$ based on our Figure 2 and thus the correction term is ignored as shown in Equation (7).  

To obtain Equation (8), we assume that the gas stream flowing under the disk is $\bf{v_{u} \approx \beta v_{o}}$ where $\beta$ is a fraction.  We also substitute for mass transfer rate $ \dot{M} = \rho$$|\bf{v_{o}}|$$A_{c}$ where $A_{c}=\pi b^{2}$ is the cross sectional area of the gas stream of radius $b$ near the bright spot.  Once again we note that, to first approximation, Equation (8) does not depend on magnetic fields or radiation sources. In addition, Equation (8) does not depend on mass.  However, the versions of lift force in Equations (7) and (8) are proportional to the surface area subject to lift, the density of the gas stream, and the gas stream speed or similarly the mass transfer rate.  

We can now make a simple comparison of the lift forces in Equations (4) and (8).  For lift to occur,  \( \bf{|F_{L_{2}}|} > \bf{|F_{L_{1}}|} \) and for lift to be maintained, \( \bf{|F_{L_{2}}|} = \bf{|F_{L_{1}}|} \).  Therefore, 
\begin{eqnarray}
|\dot{M}|  & \ge & \frac{32G \Sigma m M_{1}}{9r_{d}^{2} |v_{o}| (1-\beta^{2})}  \left( \frac{b}{r_{d}}\right)^{2}\sin\theta + \frac{3r_{d}GmM_{2}}{2|v_{o}|(1-\beta^{2})(d^{2}+\frac{9}{16}r_{d}^{2}-\frac{3}{2}r_{d}d\cos\theta)^{3/2}}  \left( \frac{b}{r_{d}}\right)^{2}\sin\theta 
\end{eqnarray}
\noindent
for the disk tilt to be initiated and/or maintained.  

Once again note that if the mass of the primary is significantly more than the mass of the secondary or the secondary is located very far from the primary, then Equation (9) reduces to the first term on the right hand side.  This scenario may be of interest to protostellar and protoplanetary systems and is of interest to our example shown in \S2.1.  In addition to the parameters introduced in \S2.1, if we assume 100,000 gas particles in the disk and \( \rho \sim 10^{-7} \) kg m$^{-3}$, then \( \Sigma m \approx 10^{-11}M_{\odot} \) (see Montgomery 2009b).  If the disk tilt is small such as five degrees, then $ |\bf{F_{L_{1}}}|$ $\approx 10^{21} $ N from Equation (4).  If we assume a supersonic velocity $\bf{|v_{o}|}$=5$\times10^{5}$ m s$^{-1} $ near the bright spot and a small difference in speeds above and below the disk ($\beta=0.9$) then $\bf{|F_{L_{2}}|}$ $\approx 10^{21} $ N from Equation (7).  The forces balance and the disk can remain tilted due to the lift force.  To find the minimum mass transfer rate to tilt the disk, we would have to know the size of the gas stream relative to the size of the disk.  If we assume $r_{d} \approx$ 100b, then \( \dot{M} \ge 5 \times 10^{12} \) kg s$^{-1}$ or equivalently \( \dot{M} \ge 2 \times 10^{-13} \) M$_{\odot}$ yr$^{-1}$.  As shown in Montgomery (2009b), all CV DN systems that are known to retrogradely precess have mass transfer rates several orders of magnitude above this calculated value.  Further, models of planet formation find minimum mass transfer rates (see e.g., Ida \& Lin 2008) that are significantly larger than the minimum gross magnitude we calculate here.  At first glance, our calculated value for minimum mass transfer rate to generate a disk tilt may seem ridiculously low.  However, as found in Montgomery (2009b), a small disk tilt would not necessarily translate into an observable negative superhump frequency.  Therefore, a higher mass transfer rate is likely needed to generate an observable signal that would indicate disk tilt.  Also note that for CV DN systems, the primary mass is that of a white dwarf, the geometry of the system is dependent on the secondary-to-primary mass ratio, and speeds near the bright spot are known to be supersonic.  As a result, the right hand side to Equation (9) should be nearly constant and thus CV DN systems that are known to retrogradely precess should nearly have the same mass transfer rate.  As shown in Montgomery (2009b), all known retrogradely precessing CV DN systems have mass transfer rates that are approximately within one order of magnitude.   

To obtain a more precise value for $\bf{v_{o}}$ in Equation (9), we would need to employ conservation of angular momentum and conservation of mechanical energy at the bright spot and at the inner Lagrange point $L_{1}$.  To find $b$, we would need to make use of the equation of continuity for fluid flow near these same points.  Also we would need to know the variation of the gas stream density and speeds over and under the disk, values that can be found from computer simulations but are best found from experiments in the laboratory.  

As a final point, we note that the right side of Equation (4) does not usually equate to the right side of Equation (7) or (8) in the laboratory.  Therefore, the ratio of the left to the right side of Equation (7) is found.  This dimensionless Coefficient of Lift is 

\begin{equation}
C_{L} = \frac{|F_{L_{2}}|}{\frac{1}{2}\rho(v_{o}^{2}-v_{u}^{2})A_{s}+ \rho |a_{G}| (r_{o}-r_{u}) }.
\end{equation}

\noindent
Typically coefficients of lift are tabulated from experiments on various geometrical shapes that are subject to a particular gas stream fluid.  The authors are not aware of such studies on accretion gas overflowing a Keplerian gas disk of the same composition, although they are likely to be important.  We assume these coefficients are small, though.  Likewise, we assume Coefficients of Drag are also small.  

\section{DISCUSSION}
\subsection{Change in Speed or Change in Direction?}
If gas stream speeds vary above and below an accretion disk, then lift can be generated.  Likewise, lift can also be generated if the gas stream flow paths vary in direction above and below the accretion disk.  As infall onto accretion disks is not likely to be symmetric, varying gas stream speeds above and below the disk is the likely source to generate disk tilt.  However, as the gas stream strikes the disk at an oblique angle and as the gas particles are traveling in the same direction and the rim of the disk is not constant in height, disk tilt is likely initiated by varying gas stream flow paths above and below the disk at the bright spot.  Because disks need to be tilted above a few degrees to be seen as modulations in light curves, initiation of disk tilt is not likely to be instantaneous.  However, once initiated above a few degrees, disk tilt results in additional changes in gas stream flow paths that induce additional disk tilt; growth could then progress rapidly.  

\subsection{Tilt or Warp?}
In this work, we assume a disk responds to lift by tilting as a unit.  We show that lift significantly depends on the difference in variation of the gas stream above and below the disk.  We assume a small difference in gas stream velocity above and below the disk to initiate a disk tilt and to maintain a disk tilt.  In this work, we did not consider larger differences in gas stream speeds above and below the disk which may cause the disk to locally warp.  Larger differences in gas stream speeds above and below the disk are expected further from the disk's rotation axis and hence may be of interest for protostellar, protoplanetary, and AGNs, for example.

\subsection{Inside-Out or Outside-In Warp?}
Some disks appear to be more warped in inner annuli (an inside-out warp) and some appear to be more warped in outer annuli (outside-in warp) and some disks appear to be entirely tilted.  Radiation and magnetic sources can warp inner annuli of a disk, for example.  This work introduces a model to warp or tilt a disk that can affect the entire disk.  However in some larger disk systems, infall onto a disk is likely to vary more further from the rotation axis of the disk, thereby indicating an outside-in warp.  In addition to radiation and magnetic sources, disk warp/tilt can likely be explained by differing gas stream speeds above and below the disk.

\subsection{Tilt by Planets?}
As protostellar clouds are not symmetric (e.g., Ciolek \& Basu, 2006), infall of gas to form a protostellar disk is unlikely to be symmetric as well.  Likewise, a protoplanetary disk is unlikely to be symmetric during planet formation.  In the disk of Beta Pictoris, Telesco et al. (2005) find a 4.6$^{o}$ disk tilt which they attribute to planet formation.  However, this work suggests that disk tilt could be from asymmetric infall during the formation of the disk.  

\subsection{Other Hydrodynamic Torques}
Hydrodynamic forces applied not at a center of mass could cause rotation of an object around its three principle cartesian (x', y', z') body axes yielding torques commonly called pitch, roll, and yaw.  In this work, we consider pitch but not roll or yaw.  In this work, we did not factor in other sources that may warp or tilt disks such as central radiation sources or magnetic fields.  In certain systems, these sources may overpower pitch effects on the disk due to lift.   

\subsection{Disk Composition}
As shown in this work, lift depends on the mass of the disk and thus we can say that lift indirectly depends on disk composition.  Gaseous, debris, and dust disks have different compositions and therefore they have different masses.  As a result, these different disks are likely to respond to lift differently.  

\subsection{Pitch Angles and Jets}
In this work, we find that an accretion disk can tilt around the line of nodes.  Because the gas stream flowing over and under the rim of the disk at the bright spot eventually rejoins the disk downstream, the density and/or speeds of the overflowing gas stream may be different above and below the disk as well as on either side of the line of nodes.  As a result, the degree of pitch around the line of nodes for each semi-circular disk wedge may not be the same (see right panel of Figure 3).  If jets are slaved to the disk, then slightly different disk pitch angles could explain slightly misaligned jets  (see e.g., Coffey, Bacciotti, \& Podio 2008).

\subsection{Magnetic Field Lines That Thread Disks}
To explain the origin of winds, jets, and bipolar outflows from young stellar objects, Shu et al. (1994) introduce threading of an accretion disk by a strong magnetic field from a stellar magneton.  The magnetic field tries to vertically thread the disk everywhere, however, the diamagnetic properties of the disk plasma prevent the vertical penetration.  Specifically, shielding currents arise in the surface layers of the disk and these currents generate a counter-magnetic field that cancels the normal component of the stellar magnetic field just above and below the disk.  Therefore, for the magnetic field to penetrate the disk, the magnetic field must be angled to the disk face.  Shu et al. (1994) suggest that inside a co-rotation radius in the disk (i.e., a radius where the disk where the star and disk co-rotate), closed magnetic field lines curve inward around the inner disk radius, allowing the accreting gas stream to attach itself to the magnetic field where it can be funneled towards the poles of the star.  Outside this co-rotation radius, open field lines curve outward.  This is the so-called X-wind model.  However, the model cannot explain all known exoplanets at their various radii.  Specifically, tidal interaction and angular momentum exchange with the central object and the creation of an inner hole at the co-rotation radius cannot reproduce all observations of exoplanet mass-periods.  Further, the origin of the disk magnetic field is still in contention, the geometry is complicated, and movement of the dead zone (i.e., disk mid-plane) may be additional concerns.  By allowing for the disk to tilt/warp due to lift, the vertical component of magnetic field lines from the central primary source can now penetrate the disk.  Combined with the natural decrease in magnetic field strength with radius from the central source, funneling from inner disk annuli may be more easily explained.  

Disk tilt might also help explain planet migration to innermost annuli.  Magnetic fields far from the central source should penetrate the disk more vertically in a non-tilted disk.  However, if the disk is tilted, then magnetic field lines from the central source can penetrate the disk at an angle.  

\subsection{Cyclic Disk Tilt}
In this work, we show that disks can tilt due to slightly different supersonic flow speeds of accretion over and under a disk that surrounds a compact central source.  If mass transfer rates decrease to the point that lift cannot be maintained, disk tilt also decreases.  If mass transfer rates were to increase at some later point in time (e.g., outburst events), disk tilt could once again be achieved.  This cyclic disk tilt could explain cyclic observations in systems suspected of disk tilt e.g., TT Ari.  

\section{SUMMARY, CONCLUSIONS, AND FUTURE WORK}
In this work we study how and why accretion disks tilt.  How accretion disks could tilt is by lift.  Why accretion disks tilt depends on disk mass, shape, and size as well as gas stream speed, density, inclination, viscosity, and compressibility.  

Gas stream overflow of the disk edge is expected because  the disk height is less than the width of the gas stream.  Because the gas stream strikes the disk at the bright spot and energy is lost to the bright spot, the speeds and density of the gas stream flowing over and under the disk may not be the same.  As the disk is a fluid, the shape of the rim is also likely to contribute to differential gas stream flow over and under the disk.  Because the period of Keplerian motion in disk annuli is faster than the period the gas stream orbits, the entire disk seems to experience the differing speeds of the overflowing gas stream.  These differing, yet supersonic, gas stream speeds and hence mass transfer rate contribute to hydrodynamic pressure which could cause lift on the disk.  Other factors also contribute to lift such as the elliptical shape of the disk and asymmetric orientation of the disk with the gas stream.  As the mass of the disk is a factor, the composition of the disk is also a factor.  We note that lift does not require radiation or magnetic sources to first approximation.  

Because the lift force acts at the center of pressure which is not at the disk's center of mass, a torque is imparted on the disk that causes the disk to pitch.  The pitch is around the disk's line of nodes, as expected.  The greater the lift force, the greater the pitch angle in the direction of lift.  The massive compact central object, which is at the center of the disk's mass and on the pitch rotation axis of the disk (i.e., line of nodes), contributes very little to the torque.  In this work, we do not study other torques on the disk that may result in disk yaw or roll.  

Because the density of the gas stream may decrease as the gas stream overflowing the disk rejoins the disk and because gas flow speeds may be different above and below the disk, pitch angels on either side of the line of nodes may not be the same.  These small differences in pitch angles may explain small differences in jet angles from the same disk.    

As higher mass transfer rates are needed to induce and sustain lift, if mass transfer rate ebbs or ceases, then so will lift.  If in a binary, tidal effects attempt to re-align the disk with the orbital plane.  If isolated, the mis-aligned angular momentum vectors of the rotating disk and the massive compact central object attempt to re-align.  The many factors that affect the growth and decay of lift could explain the observed, cyclic retrograde precessional signals that appear and disappear in systems like TT Ari.    

Typically lift and drag on various geometrical shapes in various gases are established in the lab.  Lacking this information, we provide an estimate to the predictability of lift.  To do so, we neglect viscosity and compressibility of the gas.  By making reasonable assumptions, we find that lift can be generated in e.g., non-magnetic CV DN systems.  We also establish a minimum value to mass transfer rate to generate lift, a value that is lower than all known retrogradely precessing CV DN systems and possibly all soft X-ray transients and a value that is lower than that expected for planet formation.  Because a disk tilt greater than three degrees is needed to generate a statistically significant observational signal, a mass transfer rate higher than the minimum is likely needed.  Our minimum mass transfer rate suggests that many accreting systems are likely to tilt such as neutron star, X-ray binary, CV, T Tauri, black hole and AGN systems.    

Because young protostellar and protoplanetary systems also have high mass transfer rates and infall of gas is unlikely to be symmetric, we suggest that these young accretion disks may also tilt due to lift.  By allowing for disk tilt, the vertical component of magnetic field lines from a central source can penetrate the tilted disk, simplifying the geometry which may be helpful in studies of jet formation and planet migration.  As tilt is associated with retrograde precession, we predict that these young disks should also retrogradely precess and this study is future work.  Using streamlines as predictors of accretion stream overflow is also future work.  

\section*{Acknowledgments}
This work was supported by the FP6 CONSTELLATION Marie Curie RTN which is governed by contract number MRTN-CT-2006-035890 with the European Commission.  We would like to thank mechanical engineering undergraduate Justin Hodges, a recipient of an NSF/UCF EXCEL grant, who calculated Reynolds numbers for accretion flows.  We would also like to thank undergraduate physics student Emmanuel Cruz, also a recipient of an NSF/UCF EXCEL grant, who compiled a short list of different type celestial objects that have high mass transfer rates and UCF graduate Mark Guasch who found the relative density/internal energy variations in the numerically simulated accretion disks.  We would also like to thank the anonymous referee.

}

\clearpage
\begin{figure}
\caption{Sketch of a tilted accretion disk.  The secondary of the binary is the point mass $M_{2}$ located a distance $d$ from the origin.  Solid line axes represent the orbital plane with rotation in the prograde, counter-clockwise direction around the $\bf{z}$-axis with speed $\dot{\omega}$.  The primary is at the origin and rotates counter-clockwise with speed $\dot{\epsilon}$ around the dashed $\bf{z'}$ polar axis where $\bf{(x',y',z')}$ are the body axes.  The z' axis is at an obliquity angle $\theta$ relative to the z-axis and, because of geometry, the disk is also tilted $\theta$ relative to the $\bf{y}$ axis. Note that the primary is small compared to the size of the disk and is thus hidden from view.  Forces balancing the tilted disk are labeled $\bf{F_{L}}$ and $\bf{F_{net}}$ and are located at a disk radial distance 3/4$r_{d}$. $F_{G}$ is the gravitational force between a gas particle in the disk and the primary.}
\label{Figure 1.}
\end{figure}

\begin{figure}
\caption{An artificially tilted accretion disk is shown face-on in relative density (left), and a non-tilted accretion disk is shown edge-on in relative internal energy (right).  In both panels, red indicates lowest density/internal energy whereas violet represents highest density/internal energy.  Relative densities and internal energies have the same scale:  (violet/red$\ge$83\%), (83\%$>$blue/red$\ge$29\%), (29\%$>$white/red$\ge$18\%), (18\%$>$green/red$\ge$8\%), (8\%$>$yellow/red$\ge$4\%), (4\%$>$orange/red$\ge$0.6\%), (0.6\%$>$red$\ge$0).  The figures are from  numerical simulations discussed in Montgomery (2009b).  Note the edge-on plot is a cut-away of the disk to show the locations of the innermost annuli in blue to the outermost annuli in red. The entire disk rim, colored blue and white in relative density, is red in internal energy (not shown).  }
\label{Figure 2.}
\end{figure}

\begin{figure}
\caption{
Sketch of gas flow streamlines over and under an non-tilted, edge-on accretion disk (top panel) and a tilted, edge-on accretion disk (bottom panel).  Tilt angles and thus streamlines are exaggerated to show detail.  Wider/closer spacing of streamlines indicates slower/faster flow speeds and higher/lower pressures.  The gas stream overflowing the disk edge at the bright spot eventually rejoins the disk downstream as shown.  Note that we assume in the figure that the gas speeds over/under the disk significantly vary along the gas stream path as shown by the streamlines.  The gas stream may very well remain intact in which case the streamlines above/below the disk would rejoin the disk in approximately the same location.  
 }
\label{Figure 3.}
\end{figure}

\end{document}